\documentstyle[twoside,fleqn,espcrc2]{article}

\input epsf
\def\spose#1{\hbox to 0pt{#1\hss}}
\def\lsim{\mathrel{\spose{\lower 3pt\hbox{$\mathchar"218$}}
 \raise 2.0pt\hbox{$\mathchar"13C$}}}
\def\gsim{\mathrel{\spose{\lower 3pt\hbox{$\mathchar"218$}}
 \raise 2.0pt\hbox{$\mathchar"13E$}}}

\newcommand{\AmS}{{\protect\the\textfont2
  A\kern-.1667em\lower.5ex\hbox{M}\kern-.125emS}}

\begin{document}

\begin{titlepage}

\begin{flushright}
CERN-TH/97-169\\
hep-ph/9707368
\end{flushright}

\vspace{1.5cm}

\begin{center}
\Large\bf A New Look at Hadronic B Decays
\end{center}

\vspace{1.2cm}

\begin{center}
Matthias Neubert\\
{\sl Theory Division, CERN, CH-1211 Geneva 23, Switzerland}
\end{center}

\vspace{1.3cm}

\begin{center}
{\bf Abstract:}\\[0.3cm]
\parbox{11cm}{
We present a detailed study of non-leptonic two-body decays of $B$
mesons based on a generalized factorization hypothesis. We discuss the
structure of non-factorizable corrections and present arguments in
favour of a simple phenomenological description of their effects. We
discuss tests of the factorization hypothesis and show how it may be
used to determine unknown decay constants. In particular, we obtain
$f_{D_s}=(234\pm 25)$~MeV and $f_{D_s^*}=(271\pm 33)$~MeV.}
\end{center}

\vspace{1cm}

\begin{center}
{\sl To appear in the Proceedings of the\\
High-Energy Physics Euroconference on Quantum Chromodynamics\\
Montpellier, France, 3--9 July 1997}
\end{center}

\vspace{1.5cm}

\vfil
\noindent
CERN-TH/97-169\\
July 1997

\end{titlepage}

\thispagestyle{empty}
\vbox{}
\newpage

\setcounter{page}{1}


\title{A new look at hadronic B decays}

\author{Matthias Neubert
\address{Theory Division, CERN, CH-1211 Geneva 23, Switzerland}}

\begin{abstract}
We present a detailed study of non-leptonic two-body decays of $B$
mesons based on a generalized factorization hypothesis. We discuss the
structure of non-factorizable corrections and present arguments in
favour of a simple phenomenological description of their effects. We
discuss tests of the factorization hypothesis and show how it may be
used to determine unknown decay constants. In particular, we obtain
$f_{D_s}=(234\pm 25)$~MeV and $f_{D_s^*}=(271\pm 33)$~MeV.
\end{abstract}

\maketitle

\section{INTRODUCTION}

\begin{figure}[tb]
\epsfxsize=7.5cm
\centerline{\epsffile{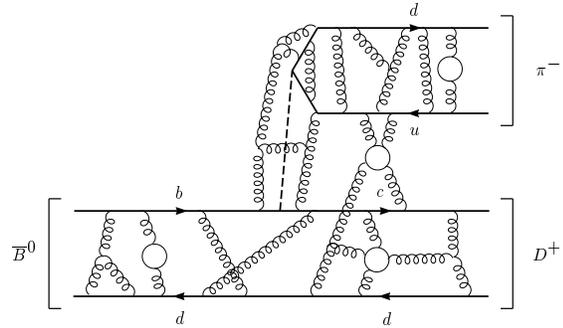}}
\vspace{-0.8cm}
\caption{\label{fig:nonlep}
Strong-interaction effects in a non-leptonic decay.}
\vspace{-0.5cm}
\end{figure}

The weak decays of hadrons containing a heavy quark offer the most
direct way to determine the weak mixing angles of the
Cabibbo-Kobayashi-Maskawa matrix and to explore the physics of CP
violation. However, an understanding of the connection between quark
and hadron properties is a necessary prerequisite for a quantitative
theoretical description of these processes. The complexity of
strong-interaction effects increases with the number of quarks
appearing in the final state. Bound-state effects in leptonic decays
can be lumped into a single parameter (the ``decay constant''), while
those in semileptonic decays are described by invariant form factors,
depending on the momentum transfer $q^2$ between the hadrons.
Approximate symmetries of the strong interactions help us to constrain
the properties of these form factors \cite{review}. For non-leptonic
decays, on the other hand, we are still lacking a comprehensive
understanding of strong-interaction effects even in simple decay modes.
The problem is exemplified in Fig.~\ref{fig:nonlep}, which shows
multiple exchanges of gluons between the quarks in the initial and
final states. The intricate interplay between weak and strong forces
has many surprising consequences, whose understanding is a challenge to
theory. Examples are the $\Delta I=\frac 12$ selection rule in $K$
decays, the difference in the lifetimes of the charm mesons $D^+$ and
$D^0$, and the difference of the lifetimes of the $\Lambda_b$ and $B$
particles. Although strong-interaction are more dramatic at low
energies, they are still hard to understand even in $B$ decays.

At tree level in the Standard Model, non-leptonic weak decays are
mediated by a single $W$-exchange diagram. When the external quarks
have energies much below the electroweak scale, this process can be
described by a local four-fermion interaction, which gets modified if
hard gluon exchanges between the quarks are included. Their effects can
be taken into account by using the renormalization group to evolve the
effective interaction from the electroweak scale down to a scale $\mu$
of order $m_b$, the mass of the decaying $b$ quark. For the case of
$b\to c\bar u d$ transitions, e.g., the relevant part of the effective
Hamiltonian is
\begin{equation}
   \frac{G_F}{\sqrt 2}\,V_{cb} V_{ud}^* \Big\{
   c_1(\mu)\,(\bar d u) (\bar c b)
   + c_2(\mu)\,(\bar c u) (\bar d b) \Big\} \,,
\end{equation}
where $(\bar d u)=\bar d\gamma^\mu(1-\gamma_5) u$ etc.\ are
left-handed, colour-singlet quark currents. The Wilson coefficients
$c_i(\mu)$ are known to next-to-leading order~\cite{ACMP,BuWe}. At the
scale $\mu=m_b$, they have the values $c_1(m_b)\approx 1.13$ and
$c_2(\mu)\approx-0.29$. These coefficients take into account the
short-distance corrections arising from the exchange of hard gluons.
The effects of soft gluons (with virtualities below the scale $\mu$)
remain in the hadronic matrix elements of the local four-quark
operators. A reliable field-theoretic calculation of these matrix
elements is the obstacle to a quantitative theory of hadronic weak
decays.

Using Fierz identities, the four-quark operators in the effective
Hamiltonian may be rewritten in various forms. It is particularly
convenient to rearrange them in such a way that the flavour quantum
numbers of one of the quark currents match those of one of the hadrons
in the final state of the considered decay process. As an example,
consider the decays $B\to D\pi$. Omitting common factors, the various
decay amplitudes may be written as
\begin{eqnarray}
   A_{B^0\to D^+\pi^-} &=& \Big( c_1 + \frac{c_2}{N_c} \Big)
    \langle D^+\pi^-|(\bar d u)(\bar c b)|B^0\rangle \nonumber\\
   &&\mbox{}+ c_2\,\langle D^+\pi^-|\textstyle\frac 12
    (\bar d t_a u)(\bar c t_a b)|B^0\rangle \,, \nonumber\\
   A_{B^0\to D^0\pi^0} &=& \Big( c_2 + \frac{c_1}{N_c} \Big)
    \langle D^0\pi^0|(\bar c u)(\bar d b)|B^0\rangle \nonumber\\
   &&\mbox{}+ c_1\,\langle D^0\pi^0|\textstyle\frac 12
    (\bar c t_a u)(\bar d t_a b)|B^0\rangle \,, \nonumber\\
   A_{B^-\to D^0\pi^-} &=& A_{B^0\to D^+\pi^-} - \sqrt 2\,
    A_{B^0\to D^0\pi^0} \,,
\label{ampl}
\end{eqnarray}
where $t_a$ are the SU(3) colour matrices. The last relation follows
from isospin symmetry of the strong interactions. The three classes of
decays shown above are referred to as class-1, class-2 and class-3,
respectively \cite{BSW}.

\section{FACTORIZATION HYPOTHESIS}

The above decay amplitudes contain the ``factorizable contributions''
\begin{eqnarray}
   {\cal F}_{(BD)\pi} &\equiv& \langle\pi^-|(\bar d u)|0\rangle\,
    \langle D^+|(\bar c b)|B^0\rangle \,, \nonumber\\
   {\cal F}_{(B\pi)D} &\equiv& \langle D^0|(\bar c u)|0\rangle\,
    \langle\pi^-|(\bar d b)|B^0\rangle \,.
\end{eqnarray}
The matrix elements in this equation are known in terms of the meson
decay constants $f_\pi$ and $f_D$, and the transition form factors for
the decays $B\to D$ and $B\to\pi$, respectively. Most of these
quantities are accessible experimentally. Of course, the matrix
elements appearing in (\ref{ampl}) also contain other, non-factorizable
contributions. In general, we may define process-dependent hadronic
parameters $\varepsilon_1$ and $\varepsilon_8$, which contain the
non-factorizable corrections, in such a way that the decay amplitudes
take the form
\begin{eqnarray}
   A(B^0\to D^+\pi^-) &=& a_1\,{\cal F}_{(BD)\pi} \,,
    \nonumber\\
   A(B^0\to D^0\pi^0) &=& a_2\,{\cal F}_{(B\pi)D} \,,
\end{eqnarray}
with \cite{Chen}--\cite{NeSt}
\begin{eqnarray}
   a_1 &=& \left(c_1(\mu) + \frac{c_2(\mu)}{N_c}\right)
    \left[ 1 + \varepsilon_1^{(BD)\pi}(\mu) \right] \nonumber\\
   &&\mbox{}+ c_2(\mu)\,\varepsilon_8^{(BD)\pi}(\mu) \,, \nonumber\\
   a_2 &=& \left(c_2(\mu) + \frac{c_1(\mu)}{N_c}\right)
    \left[ 1 + \varepsilon_1^{(B\pi)D}(\mu) \right] \nonumber\\
   &&\mbox{}+ c_1(\mu)\,\varepsilon_8^{(B\pi)D}(\mu) \,.
\label{a1a2}
\end{eqnarray}
We stress that these expressions are exact; they are just a way to
parametrize the relevant matrix elements of four-quark operators. The
effective coefficients $a_i$ take into account all contributions to the
matrix elements and are thus $\mu$ independent. The scale dependence of
the Wilson coefficients is exactly balanced by that of the hadronic
parameters.

Additional insight can be gained by combining these results with the
$1/N_c$ expansion \cite{BGR}. At the scale $\mu=O(m_b)$, the
large-$N_c$ counting rules of QCD imply $c_1=1+O(1/N_c^2)$,
$c_2=O(1/N_c)$, $\varepsilon_1=O(1/N_c^2)$, and
$\varepsilon_8=O(1/N_c)$. (For scales much lower than $m_b$, the
counting rules for the Wilson coefficients $c_i(\mu)$ are spoiled by
large logarithms.) Hence, we expect that $|\varepsilon_1|\ll 1$,
whereas contributions from $\varepsilon_8$ can be larger. Using these
results, we find from (\ref{a1a2})
\begin{eqnarray}
   a_1 &=& c_1(m_b) + O(1/N_c^2) \,, \nonumber\\
   a_2 &=& c_2(\mu) + c_1(\mu)\!\left(\! \frac{1}{N_c}
    + \varepsilon_8^{(B\pi)D}(\mu) \!\right) + O(1/N_c^3) \nonumber\\
   &\equiv& c_2(m_b) + \zeta\,c_1(m_b) \,,
\label{xidef}
\end{eqnarray}
where $\zeta=1/N_c+\varepsilon_8(m_b)$ is a process-dependent hadronic
parameter of order $1/N_c$ \cite{NeSt}. It is important to stress that
the naive choice $a_1=c_1+c_2/N_c$ and $a_2=c_2+c_1/N_c$, which is
often referred to as ``factorization hypothesis'', does not correspond
to {\em any} consistent limit of QCD; in particular, this is not a
prediction of the $1/N_c$ expansion. The more general expression for
$a_2$ given above was first introduced in Ref.~\cite{BSW}. Since the
parameter $\varepsilon_8$ is of order $1/N_c$, the two contributions to
$\zeta$ are of the same magnitude, and hence $\zeta$ should be
considered as an unknown dynamical parameter. As a general rule, we
expect that non-factorizable corrections are small in class-1
transitions. In class-2 decays, on the other hand, the contribution
proportional to $\varepsilon_8$ is enhanced by the large value of the
ratio $c_1/c_2=O(N_c)$, and non-factorizable contributions can
therefore be sizeable.

\begin{figure}
\epsfxsize=7.5cm
\centerline{\epsffile{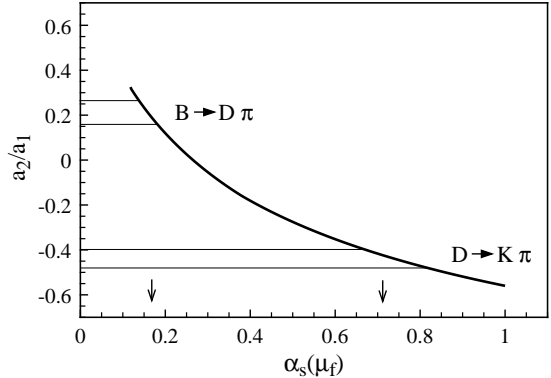}}
\vspace{-0.8cm}
\caption{\label{fig:a2a1}
The ratio $a_2/a_1$ as a function of the running coupling constant
evaluated at the factorization scale. The bands indicate the
phenomenological values of $a_2/a_1$ extracted from $B\to D\pi$ and
$D\to K\pi$ decays.}
\vspace{-0.5cm}
\end{figure}

The parameter $\varepsilon_8$ obeys the renormalization-group equation
\begin{equation}
   \mu\,\frac{d}{d\mu}\,\varepsilon_8^P(\mu)
   \approx -\frac{4\alpha_s}{3\pi} \,,
\label{RGE}
\end{equation}
where the superscript $P$ represents the dependence on the decay
process. Let us assume that, for each process, there exists a
``factorization scale'' $\mu_f$ such that $\varepsilon_8^P(\mu_f)=0$.
(We will see later that this is indeed the case for two-body decays of
$B$ mesons.) Then (\ref{RGE}) implies that the phenomenological
parameter $\zeta$ is given by \cite{NeSt}
\begin{equation}
   \zeta \approx \frac{1}{N_c} - \frac{4\alpha_s}{3\pi}\,
   \ln\frac{m_b}{\mu_f} \,.
\end{equation}
Based on the colour transparency argument of Bjorken \cite{Bj89}, we
expect that $\mu_f$ scales with the energy of the outgoing hadrons in a
decay process. A fast-moving pair of quarks in a colour-singlet state
acts as a colour dipole and decouples from soft gluons. Only hard
gluons, with virtualities of order the energy of the outgoing
particles, can rearrange the quarks and thus spoil factorization. On a
qualitative level, the connection between the factorization scale and
the energy release in the final state can be seen from
Fig.~\ref{fig:a2a1}, where we show the ratio $a_2/a_1$ as a function of
$\alpha_s(\mu_f)$. As we will see later, the value preferred by $B\to
D\pi$ decays is positive and corresponds to a rather small coupling,
indicating $\mu_f=O(m_b)$ for these processes. On the other hand, $D$
decays indicate a negative value of $a_2/a_1$, corresponding to a lower
value of the factorization scale. This is in accordance with the fact
that in these processes the energy released to the final-state
particles is much smaller.

In two-body decays of $B$ mesons, the energy release (per particle) in
different processes differs by less than about 1~GeV. Combining this
observation with the empirical fact that in $B\to D\pi$ decays the
scale $\mu_f$ is of order $m_b$, we find that the changes of the
phenomenological parameter $\zeta$ in different decay channels are of
order \cite{NeSt}
\begin{equation}
   \Delta\zeta \sim \frac{4\alpha_s}{3\pi}\,\frac{\Delta\mu_f}{m_b}
   \sim \mbox{few \%} \,.
\end{equation}
Thus, on rather general grounds, we expect that in two-body $B$ decays,
to a good approximation, $\zeta$ is process independent. We shall refer
to this assumption, together with the prediction for the parameters
$a_1$ and $a_2$ given in (\ref{xidef}), as {\em generalized
factorization
hypothesis}.

\section{TESTS AND PREDICTIONS}

Adopting the theoretical framework described above, a prediction of
hadronic decay amplitudes needs as input information meson decay
constants and form factors. The decay constants of many light mesons
are known with high accuracy from hadronic $\tau$ decays, and from the
electromagnetic decays of vector mesons \cite{NeSt}. The meson form
factors of quark currents can, to some extend, be extracted from
semileptonic decays. Whereas the various $B\to D^{(*)}$ transition form
factors are well-known by combining data from semileptonic decays with
heavy-quark symmetry relations \cite{review}, the available information
about form factors describing the decays of $B$ mesons into light
mesons is largely model dependent. As a consequence, for all class-1
decays considered below the factorized decay amplitudes can be
predicted without any model assumptions; however, the theoretical
predictions for class-2 amplitudes involve larger theoretical
uncertainties. In this work, we will adopt the NRSX model \cite{NRSX}
to calculate the meson form factors. Because of lack of information, we
shall neglect final-state interactions between the produced hadrons in
the final state. In energetic two-body decays of $B$ mesons, these
effects are expected to be small \cite{NeSt,Pham}.

Our predictions for the branching ratios of some of the dominant
non-leptonic two-body decays of $B$ mesons are given in
Table~\ref{tab:Bdecays}. The QCD coefficients $a_1$ and $a_2$, as well
as the (poorly known) decay constants of charm mesons, have been left
as parameters in the theoretical expressions. For comparison, we show
the world average experimental results for the branching ratios, as
recently compiled in the review article in Ref.~\cite{newrev}.

\begin{table*}[p]
\setlength{\tabcolsep}{1.5pc}
\newlength{\digitwidth} \settowidth{\digitwidth}{\rm 0}
\catcode`?=\active \def?{\kern\digitwidth}
\caption{Theoretical predictions for the branching ratios (in \%) of
non-leptonic $B$ decays. In the third column, the factors containing
not well known decay constants are suppressed.}
\label{tab:Bdecays}
\begin{tabular*}{\textwidth}{lccc}
\hline \\
$B^0$ Modes & NRSX Model & $a_1=1.08$ & Experiment \\
 & & $a_2=0.21$ & \\
\hline
Class-1 & & & \\
\hline
$D^+\pi^-$ & 0.257 $a_1^2$ & 0.30 & $0.31\pm 0.04\pm 0.02$ \\
$D^+\varrho^-$ & 0.643 $a_1^2$ & 0.75 & $0.84\pm 0.16\pm 0.07$ \\
$D^{*+}\pi^-$ & 0.247 $a_1^2$ & 0.29 & $0.28\pm 0.04\pm 0.01$ \\
$D^{*+}\varrho^-$ & 0.727 $a_1^2$ & 0.85 & $0.73\pm 0.15\pm 0.03$ \\
$D^+ D_s^-$ & 0.879 $a_1^2\,(f_{D_s}/240)^2$ & 1.03 &
 $0.74\pm 0.22\pm 0.18$ \\
$D^+ D^{*-}_s$ & 0.817 $a_1^2\,(f_{D_s^*}/275)^2$ & 0.95 &
 $1.14\pm 0.42\pm 0.28$ \\
$D^{*+} D_s^-$ & 0.597 $a_1^2\,(f_{D_s}/240)^2$ & 0.70 &
 $0.94\pm 0.24\pm 0.23$ \\
$D^{*+} D_s^{*-}$ & 2.097 $a_1^2\,(f_{D_s^*}/275)^2$ & 2.45 &
 $2.00\pm 0.54\pm 0.49$ \\
\hline
Class-2 & & & \\
\hline
$\bar K^0 J/\psi$ & 2.262 $a_2^2$ & 0.10 & $0.075\pm 0.021$ \\
$\bar K^0 \psi(\mbox{2S})$ & 1.051 $a_2^2$ & 0.05 & $<0.08$ \\
$\bar K^{*0} J/\psi$ & 3.645 $a_2^2$ & 0.16 & $0.153\pm 0.028$ \\
$\bar K^{*0} \psi(\mbox{2S})$ & 1.939 $a_2^2$ & 0.09 &
 $0.151\pm 0.091$ \\
\hline \\
$B^-$ Modes & NRSX Model & $a_1=1.08$ & Experiment \\
 & & $a_2=0.21$ & \\
\hline
Class-1 & & & \\
\hline
$D^0 D_s^-$ & 0.938 $a_1^2\,(f_{D_s}/240)^2$ & 1.09 &
 $1.36\pm 0.28\pm 0.33$ \\
$D^0 D^{*-}_s$ & 0.873 $a_1^2\,(f_{D_s^*}/275)^2$ & 1.02 &
 $0.94\pm 0.31\pm 0.23$ \\
$D^{*0} D_s^-$ & 0.639 $a_1^2\,(f_{D_s}/240)^2$ & 0.75 &
 $1.18\pm 0.36\pm 0.29$ \\
$D^{*0} D^{*-}_s$ & 2.235 $a_1^2\,(f_{D_s^*}/275)^2$ & 2.61 &
 $2.70\pm 0.81\pm 0.66$ \\
\hline
Class-2 & & & \\
\hline
$K^- J/\psi$ & 2.411 $a_2^2$ & 0.11 & $0.102\pm 0.014$ \\
$K^-\psi(\mbox{2S})$ & 1.122 $a_2^2$ & 0.05 & $0.070\pm 0.024$ \\
$K^{*-} J/\psi$ & 3.886 $a_2^2$ & 0.17 & $0.174\pm 0.047$ \\
$K^{*-} \psi(\mbox{2S})$ & 2.070 $a_2^2$ & 0.09 & $<0.30$ \\
\hline
Class-3 & & & \\
\hline
$D^0\pi^-$ & $0.274\left[a_1+1.127\,a_2\,(f_D/200)\right]^2$ &
 0.48 & $0.50\pm 0.05\pm 0.02$ \\
$D^0\varrho^-$ & $0.686\left[a_1+0.587\,a_2\,(f_D/200)\right]^2$ &
 0.99 & $1.37\pm 0.18\pm 0.05$ \\
$D^{*0}\pi^-$ & $0.264\left[a_1+1.361\,a_2\,(f_{D^*}/230)\right]^2$ &
 0.49 & $0.52\pm 0.08\pm 0.02$ \\
$D^{*0}\varrho^-$ & $0.775\,[a_1^2+0.661\,a_2^2\,(f_{D^*}/230)^2$ &
 1.19 & $1.51\pm 0.30\pm 0.02$ \\
 & $\phantom{0.775[} \mbox{}+ 1.518\,a_1 a_2\,(f_{D^*}/230)]$ &
 & \\
\hline
\end{tabular*}
\end{table*}

\boldmath
\subsection{Extractions of $a_1$}
\unboldmath

There are various ways in which to test the generalized factorization
hypothesis. The most direct one relies on the close relationship
between semileptonic and factorized hadronic decay amplitudes.
Comparing the non-leptonic decay rates with the corresponding
differential semileptonic decay rates evaluated at the same value of
$q^2$ provides a direct test of the factorization hypothesis
\cite{Bj89}. We have
\begin{eqnarray}
   \frac{\Gamma(B^0\to D^{(*)+} h^-)}
   {d\Gamma(B^0\!\to\! D^{(*)}\ell\,\bar\nu)/dq^2\Big|_{q^2=m_h^2}}
   = 6\pi^2 |V_{ud}|^2 f_h^2 a_1^2 \,, \nonumber\\
\end{eqnarray}
where $h$ denotes a light meson ($h=\pi$ or $\varrho$), and $f_h$ its
decay constant. To determine this ratio experimentally, one needs the
values of the differential semileptonic decay rate at various values of
$q^2$. They have been determined for $B\to D^*\ell\,\bar\nu$ decays in
Ref.~\cite{newrev}, using a fit to experimental data. A comparison of
the theoretical prediction with experimental data yields $a_1=1.11\pm
0.10$ for $h=\pi$, and $a_1=1.09\pm 0.13$ for $h=\varrho$.

A more precise determination of the parameter $a_1$ is obtained by
comparing the theoretical predictions for the decays $B^0\to D^{(*)+}
h^-$ (with $h=\pi$ or $\varrho$) directly with the data. For these
processes, the transition form factors and decay constants are well
known experimentally. Therefore, the theoretical uncertainties are
minimal. From a fit to the data, we obtain $a_1=1.08\pm 0.04$.

An interesting alternative extraction of $a_1$ is obtained from the
class of decays $B^0\to D^{(*)+} D_s^{(*)-}$, which involve the same
$B\to D^{(*)}$ transition form factors, but have a quite different
kinematics since there is less energy release to the final-state
particles. Unfortunately, the decay constants of $D_s$ mesons are not
well known experimentally. From a fit to the data, we find $a_1=1.10\pm
0.07\pm 0.17$, where the last error takes into account the uncertainty
in the decay constants.

The different determinations of the parameter $a_1$ agree well with
each other and yield a result that confirms the theoretical expectation
that $a_1\approx c_1(m_b)\approx 1.13$. Within the experimental errors,
there is no evidence for a process dependence of $a_1$.

\boldmath
\subsection{Extractions of $a_2$}
\unboldmath

A determination of the parameter $a_2$, as well as of the relative sign
between $a_2$ and $a_1$, is obtained by comparing the theoretical
predictions for the decays $B^-\to D^{(*)0} h^-$ (with $h=\pi$ or
$\varrho$) with the data. Since the contributions to the decay
amplitudes proportional to $a_2$ involve the $B\to h$ form factors,
they cannot be predicted in a model-independent way. Using the NRSX
model for these form factors and assigning a conservative error, we
find $a_2/a_1=0.21\pm 0.05\pm 0.04$, where the second error accounts
for the model dependence. Combining this with the value for $a_1$
determined above gives $a_2=0.23\pm 0.05\pm 0.04$. The fact that the
ratio $a_2/a_1$ is positive in $B$ decays implies that class-1 and
class-2 decay amplitudes interfere constructively. This is in contrast
with the situation encountered in charm decays, where a similar
analysis yields $a_1=1.10\pm 0.05$ and $a_2=-0.49\pm 0.04$ \cite{NRSX},
indicating a strong destructive interference. Since most $D$ decays are
(quasi) two-body transitions, this effect is responsible for the
observed lifetime difference between $D^+$ and $D^0$ mesons:
$\tau(D^+)>\tau(D^0)$. In $B$ decays, on the other hand, the majority
of transitions proceeds into multi-body final states, and moreover
there are many $B^-$ decays (such involving two charm quarks in the
final state) where no interference can occur. The relevant scale for
multi-body decay modes may be significantly lower than $m_b$, leading
to destructive interference (see Fig.~\ref{fig:a2a1}). Therefore, the
observed constructive interference in the two-body modes is not in
conflict with the fact that $\tau(B^-)>\tau(B^0)$.

An alternative determination of the magnitude (but not the sign) of
$a_2$ can be obtained from the class of decays $B\to K^{(*)}
\psi^{(\prime)}$, which are characterized by a quite different decay
kinematics. We find $a_2=0.21\pm 0.01\pm 0.04$. A comparison of this
result with the value of $a_2$ determined above provides an interesting
test of our theoretical prediction that even in decay modes with
different energy release the process dependence of $a_2$ is expected to
be mild. Within errors, there is indeed no evidence for any process
dependence. Hence, the data fully support the generalized factorization
hypothesis. From the result for $a_2$ we may then extract the value of
the phenomenological parameter $\zeta$ or, equivalently, of the
colour-octet matrix element $\varepsilon_8$. We obtain
\begin{equation}
   \zeta = 0.45\pm 0.05 \,,\quad \varepsilon_8(m_b) = 0.12\pm 0.05 \,.
\end{equation}
It would be most interesting to derive these results from a rigorous,
field-theoretical evaluation of the four-quark operator matrix
elements.

\subsection{Determination of decay constants}

As an application, we shall employ the generalized factorization
hypothesis to obtain rather precise values for the decay constants of
the $D_s$ and $D_s^*$ mesons. To this end, we derive from
Table~\ref{tab:Bdecays} the theoretical predictions for the following
ratios of decay rates:
\begin{eqnarray}
   \frac{\Gamma(B^0\to D^+ D_s^-)}{\Gamma(B^0\to D^+\pi^-)}
   &=& 1.01\,\left( \frac{f_{D_s}}{f_\pi} \right)^2 \,, \nonumber\\
   \frac{\Gamma(B^0\to D^{*+} D_s^-)}{\Gamma(B^0\to D^{*+}\pi^-)}
   &=& 0.72\,\left( \frac{f_{D_s}}{f_\pi} \right)^2 \,, \nonumber\\
   \frac{\Gamma(B^0\to D^+ D_s^{*-})}{\Gamma(B^0\to D^+\varrho^-)}
   &=& 0.74\,\left( \frac{f_{D_s^*}}{f_\varrho} \right)^2 \,,
    \nonumber\\
   \frac{\Gamma(B^0\to D^{*+} D_s^{*-})}
    {\Gamma(B^0\to D^{*+}\varrho^-)} &=& 1.68\,\left(
    \frac{f_{D_s^*}}{f_\varrho} \right)^2 \,.
\end{eqnarray}
Theoretically, these predictions are rather clean for the following
reasons: first, all decays involve class-1 transitions, so that
deviations from factorization are probably very small; secondly, the
parameter $a_1$ cancels in the ratios; thirdly, the two processes in
each ratio have a similar kinematics, so that the corresponding decay
rates are sensitive to the same form factors, however evaluated at
different $q^2$ values; finally, we may hope that also some of the
experimental systematic errors cancel in the ratios (however, we do not
assume this in quoting errors below). Combining these predictions with
the average experimental branching ratios \cite{newrev},
\begin{eqnarray}
   {\rm B}(B\to D D_s^-) &=& (0.95\pm 0.24)\% \,, \nonumber\\
   {\rm B}(B\to D D_s^{*-}) &=& (1.00\pm 0.30)\% \,, \nonumber\\
   {\rm B}(B\to D^* D_s^-) &=& (1.03\pm 0.27)\% \,, \nonumber\\
   {\rm B}(B\to D^* D_s^{*-}) &=& (2.26\pm 0.60)\% \,,
\end{eqnarray}
we find the rather accurate values
\begin{eqnarray}
   f_{D_s} &=& (234\pm 25)~{\rm MeV} \,, \nonumber\\
   f_{D_s^*} &=& (271\pm 33)~{\rm MeV} \,.
\end{eqnarray}
The result for $f_{D_s}$ is in excellent agreement with the value
$f_{D_s}=241\pm 37$~MeV extracted from the leptonic decay
$D_s\to\mu^+\nu$ \cite{Rich}. The ratio $f_{D_s^*}/f_{D_s}=1.16\pm
0.19$, which cannot be determined from leptonic decays, is in good
agreement with theoretical expectations~\cite{matthias,abada}. Finally,
we note that, assuming SU(3) breaking effects of order 10--20\%, the
established value of $f_{D_s}$ implies $f_D\approx 200$~MeV, which is
larger than most theoretical predictions.

\section{SUMMARY}

Exclusive hadronic decays of $B$ mesons are strongly influenced by the
long-range QCD colour forces. Theoretically, their description involves
hadronic matrix elements of local four-quark operators, which are
notoriously difficult to calculate. The factorization approximation is
used to relate these matrix elements to products of current matrix
elements. Conventionally, the factorized decay amplitudes depend on two
phenomenological parameters $a_1$ and $a_2$, which are connected with
the Wilson coefficients $c_i(\mu)$ appearing in the effective weak
Hamiltonian. We have shown that this approach can be generalized in a
natural way to include the dominant non-factorizable contributions to
the decay amplitudes. (The situation is more complicated in $B$ decays
into two vector mesons, see Ref.~\cite{NeSt}.) In the generalized
factorization scheme, the effective parameters $a_1$ and $a_2$ become
process-dependent. However, using the large-$N_c$ counting rules of
QCD, we have argued that in energetic two-body decays of $B$ mesons
$a_1\approx c_1(m_b)$ and $a_2\approx c_2(m_b)+\zeta\,c_1(m_b)$, where
$\zeta=O(1/N_c)$ is a dynamical parameter. Moreover, the colour
transparency argument suggests that the process dependence of $\zeta$
is likely to be very mild, so that it can be taken to be a constant for
a wide class of two-body decays. These theoretical expectations are
fully supported by the data. From a fit to the world average branching
ratios of two-body decay modes, we obtain $a_1\approx 1.08\pm 0.04$ and
$a_2\approx 0.21\pm 0.05$, corresponding to $\zeta\approx 0.45\pm
0.05$. There is no evidence for a process dependence of these
parameters; in particular, the values obtained for $a_2$ from the
decays $B\to\bar K^{(*)}\psi^{(\prime)}$ and $B^-\to D^{(*)0} h^-$,
where $h=\pi$ or $\varrho$, are in good agreement with each other.

We have discussed various tests of the generalized factorization
hypothesis by considering ratios of decay rates, and by comparing
non-leptonic decay rates with semileptonic rates evaluated at the same
value of $q^2$. Within the present experimental uncertainties, there
are no indications for any deviations from the factorization scheme in
which $a_1$ and $a_2$ are treated as process-independent hadronic
parameters. Accepting that this scheme provides a useful
phenomenological concept, exclusive two-body decays of $B$ mesons offer
a unique opportunity to measure the decay constants of some light or
charm mesons, such as the $a_1$, $D_s$ and $D_s^*$.

\section*{Acknowledgments}

The work presented here has been done in a most enjoyable collaboration
with B.~Stech, which is gratefully acknowledged.

\end{document}